\pdfoutput=1
\documentclass{article}

\usepackage{arxiv}
\usepackage{amsmath}
\usepackage[utf8]{inputenc} 
\usepackage[T1]{fontenc}    
\usepackage{hyperref}       
\usepackage{url}            
\usepackage{booktabs}       
\usepackage{amsfonts}       
\usepackage{nicefrac}       
\usepackage{microtype}      
\usepackage{lipsum}
\usepackage{graphicx}
\usepackage{subcaption}
\usepackage{float}
\graphicspath{ {./images/} }
\usepackage[round]{natbib}   
\usepackage{natbib}
\setcitestyle{aysep={}}

\title{On the relationship between the Wasserstein distance and differences in life expectancy at birth}

\author{
 Markus Sauerberg \\
  Cancer Registry Hamburg\\
  Hamburg, Germany\\
  \texttt{sauerbergmarkus@gmail.com} \\
}

\begin{document}
\maketitle
\begin{abstract}
The Wasserstein distance is a metric for assessing distributional differences. The measure originates in optimal transport theory and can be interpreted as the minimal cost of transforming one distribution into another. In this paper, the Wasserstein distance is applied to life table age-at-death distributions. The main finding is that, under certain conditions, the Wasserstein distance between two age-at-death distributions equals the corresponding gap in life expectancy at birth ($e_0$). More specifically, the paper shows mathematically and empirically that this equivalence holds whenever the survivorship functions do not cross. For example, this applies when comparing mortality between women and men from 1990 to 2020 using data from the Human Mortality Database. In such cases, the gap in $e_0$ reflects not only a difference in mean ages at death but can also be interpreted directly as a measure of distributional difference.
\end{abstract}


\section{Introduction}

The Wasserstein distance, also known as the Earth Mover’s distance, measures the distance between two probability distributions. The metric is often illustrated by imagining each distribution as a pile of dirt, where the Wasserstein distance corresponds to the minimal cost of transporting one pile into the other (see \href{https://docs.scipy.org/doc/scipy/reference/generated/scipy.stats.wasserstein_distance.html}{docs.scipy.org}). The measure originates in optimal transport theory but is now widely applied in computer science, for example in image retrieval \citep{Rubner2000} and machine learning \citep{Peyre2019}. A historical overview of the optimal transport problem can be found in \citep{Santambrogio2015}. In short, the problem of finding the optimal transport map to transform one distribution into another was first formulated in 1781 by the French mathematician Gaspard Monge. However, Monge’s original formulation remained mathematically intractable, since it was not clear whether a minimizer (an optimal transport map) always exists. In the 1940s, the Soviet mathematician Leonid Kantorovich reformulated the problem in terms of linear programming, which made it mathematically tractable and led to rigorous solutions. Since Monge and Kantorovich played central roles, the underlying problem is often referred to as the Monge–Kantorovich problem. The metric itself, however, is usually called the Wasserstein distance, after Leonid Vaserstein, who studied these distances in the 1960s \citep{Santambrogio2015}.

Using the Kantrovich formulation, the distance between two probability distributions $P$ and $Q$ can be defined as,

\begin{equation}\label{eq1}
    W_p(P,Q) = \left(\inf_{J \in \mathcal{J}(P,Q)} \int \|x-y\|^p dJ(x,y)\right)^{1/p},
\end{equation}
where $\mathcal{J}(P,Q)$ denotes a set of all couplings (also called transport plans) between $P$ and $Q$. The $\inf_{J \in \mathcal{J}(P,Q)}$ denotes the infimum, i.e., among all transport plans, we are looking for the one with the smallest cost. For each pair, $(x,y)$, the cost of moving mass from $x$ to $y$ is given by $\|x-y\|^p$. Accordingly, $\|x-y\|^p$ defines how the distance between $x$ and $y$ is measured. The outer exponent $1/p$ ensures that the resulting distance has the same units as the underlying metric. 

When $p=1$ the measure is called the Earth Mover’s Distance (EMD) and represents the minimum average absolute distance you must move mass to transform $P$ into $Q$ \citep{Wasserman2019}.

The equation \ref{eq1} gives a general formulation of the Wasserstein distance. In the one-dimensional case, the optimal transport problem is much simpler. Let $F_P$ and $F_Q$ be the cumulative distribution functions (CDF) of $P$ and $Q$, respectively. Then, the inverse of $F_P$ and $F_Q$ give the quantile functions $F_P^{-1}$ and $F_Q^{-1}$. The Wasserstein distance in the one-dimensional case, can be expressed as,

\begin{equation}\label{eq1a}
    W_p(P,Q)=\left(\int_0^1\left|F_P^{-1}(x) - F_Q^{-1}(x)\right|^p\,dx\right)^{1/p}.
\end{equation}

Further, it has been shown that in a special case - when considering distributions from a one-dimensional space and using the absolute difference, $p=1$, as the cost function, the Wasserstein distance simplifies to,

\begin{equation}\label{eq2}
    W_1(P,Q) = \int_{-\infty}^{+\infty} |F_P(x) - F_Q(x)|dx.
\end{equation}
Please refer to \cite{Santambrogio2015}, chapter 2 for more details on the equivalence of both equations.

The equation \ref{eq2} is key for assessing distributional differences between two life table age-at-death distributions. Using $p=1$ is appropriate when comparing age-at-death distributions because it treats transport costs linearly, i.e., moving probability mass from age ten to age zero indicates a cost equal to the distance between the two ages $(|x-y|=|10-0|=10)$. In other words, with $p=1$, the distance is measured simply as the absolute difference in ages. If we were to use $p=2$ instead, the cost of transporting mass grows quadratically since the distance term becomes $(|x-y|^2 = (|10-0|)^2=100)$. This quadratic scaling makes transports over long distances disproportionately expensive, e.g., between very old ages and very young ages. In contrast, the linear $p=1$ formulation preserves a direct interpretation of transport distance as the absolute age difference, which is often more meaningful for comparing age-at-death distributions.

In the same one-dimensional setting with linear cost $|x-y|$, the Wasserstein distance can equivalently be expressed using the optimal transport map. In this case, the optimal transport problem uses a Monge solution and the optimal transport map $T:\mathbb{R}\to\mathbb{R}$ is given by applying the quantile transformation,
\begin{equation}\label{eq:transport_map}
    T(x) = F_Q^{-1}\!\left(F_P(x)\right).
\end{equation}

Using this map, the $W_1$ distance can be written as,
\begin{equation}\label{monge_map}
    W_1(P,Q) = \int_{-\infty}^{+\infty} |x - T(x)| \, dP(x),
\end{equation}
which represents the expected absolute move of probability mass under the optimal transport from $P$ to $Q$ \citep[p. 32]{Peyre2019} .

To the best of my knowledge, the first scholars who applied the optimal transport theory framework to the field of demography are \cite{Oeppen2021},  \cite{Cilek2023}, and \cite{Shang2025}. While \cite{Oeppen2021} discussed optimisation in life-saving models of mortality, \cite{Cilek2023} used the Wasserstein distance to investigate cause-specific mortality differences between border regions and \cite{Shang2025} introduced an approach for forcasting age-at death distributions on the basis of CDFs which is related to the $W_1$ Wasserstein distance through equation \ref{eq1a} and \ref{eq2}.

\begin{figure}[H]
\centering
\caption{Toy example for demonstrating the idea behind the Wasserstein distance using all-cause specific mortality}
\includegraphics[width=10cm]{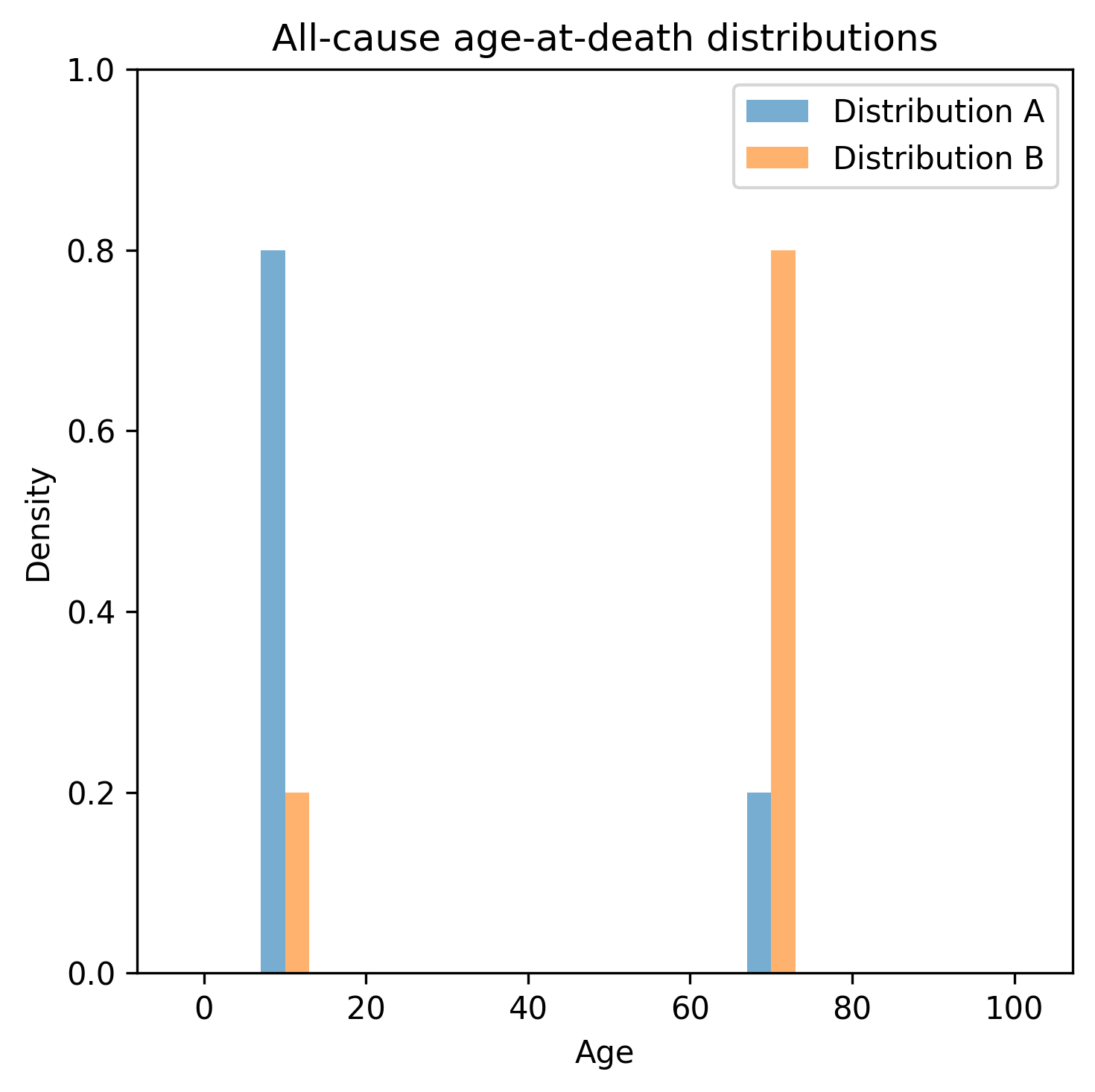}
\label{fig1d}

\medskip
\small
\begin{minipage}{\linewidth}
\raggedright
To make both distributions look the same, we need to move $0.6$ mass from age $70$ to age $10$. Hence, the Wasserstein distance is given by
$0.6 \cdot (|70-10|)=10$.
\end{minipage}

\vspace{-10pt}
\end{figure}

\section{Methods}

\section{Relationship to differences in life expectancy at birth}\label{Relationship}
Let $d_A$ and $d_B$ be two life table age-at-death distributions with corresponding survivorship functions, $l_A$ and $l_B$. Further, assume that in both life tables the radix is set to one, $l_A(0)=l_B(0)=1$, and have $l_A(\omega)=l_B(\omega)=0$ because the probability of dying at the last-open age interval, $\omega$, is one.

If the survivorship functions do not crossover,
\begin{equation}\label{eq3}
l_A(x)\ge l_B(x)\quad\forall x\in[0,\omega],
\end{equation}
then the $W_1$ Wasserstein distance between $d_A$ and $d_B$ equals the difference in life expectancy at birth,
\begin{equation}\label{eq4}
W_1(d_A,d_B)=e_{0,A}-e_{0,B}.
\end{equation}

\subsection{Proof}\label{Proof}
With a radix of one, the life table age-at-death distribution can be seen as a probability density function (PDF), and $l(x)$ gives the survival function $S(x)$. Consequently, the respective CDF is given by, 
\begin{equation}
F(x)=1-l(x).    
\end{equation}
Life expectancy at birth can be expressed as,
\begin{equation}\label{eq5}
e_0=\int_0^\omega l(x) dx.    
\end{equation}
Hence,
\begin{equation}\label{eq6}
e_{0,A}-e_{0,B}=\int_0^\omega \bigl(l_A(x)-l_B(x)\bigr)dx.
\end{equation}
As shown above, in equation \ref{eq2}, the $W_1$ Wasserstein distance, is defined as,
\begin{equation}\label{eq7}
W_1(d_A,d_B)=\int_{0}^{\omega}\bigl|F_A(x)-F_B(x)\bigr|dx.
\end{equation}

Since $F_A(x)=1-l_A(x)$ and $F_B(x)=1-l_B(x)$ we have,
\begin{equation}\label{eq8}
\bigl|F_A(x)-F_B(x)\bigr| = \bigl|(1-l_A(x))-(1-l_B(x))\bigr| = \bigl|l_A(x)-l_B(x)\bigr|.
\end{equation}
Under the condition that both survivorship functions do not crossover, equation \ref{eq3}, the integrand is non-negative. So, the absolute value can be removed,
\begin{equation}\label{eq9}
W_1(d_A,d_B)=\int_0^\omega \bigl|l_A(x)-l_B(x)\bigr|\,dx = \int_0^\omega \bigl(l_A(x)-l_B(x)\bigr)\,dx.
\end{equation}
Substituting the right-hand-side of equation \ref{eq9} into equation \ref{eq6} yields the difference in life expectancy at birth. $\square$

\subsection{Life expectancy at birth as the mean of the age-at-death distribution}\label{Equivalence}

Instead of expressing life expectancy at birth as the area under the survivorship function,
\begin{equation}\label{eq:e0_surv}
e_0 = \int_0^{\omega} l(x)dx,
\end{equation}
we can also define $e_0$ as the mean of the age-at-death distribution. To see the equivalence, we apply integration by parts to \ref{eq:e0_surv},
\begin{align}
\int_0^{\omega} l(x)dx
&= \int_0^{\omega} l(x)\cdot 1dx \\
&= \bigl[x\,l(x)\bigr]_0^{\omega} - \int_0^\omega x\,l'(x)dx \\
&= 0 - \int_0^\omega x\,l'(x)dx \\
&= \int_0^\omega x\,\bigl(-l'(x)\bigr)dx.
\end{align}
By definition, the age-at-death distribution is,
\begin{equation}
d(x) = -l'(x).
\end{equation}
For a life table with a radix of one, $\int_0^\omega d(x)dx=1$. Hence, life expectancy at birth is given by,
\begin{equation}
e_0 = \int_0^{\omega} x\,d(x)dx.
\end{equation}
More generally, normalization yields,
\begin{equation}
e_0 = \frac{\int_0^{\omega} x\,d(x)dx}{\int_0^{\omega} d(x)dx}.
\end{equation}
Thus, life expectancy at birth can be written either as the area under the survivorship function or as the mean of the age-at-death distribution. This implies that the $W_1$ Wasserstein distance between two age-at-death distributions equals the difference in their respective means, whenever the corresponding survivorship functions do not crossover,
\begin{equation}
    W_1(d_A,d_B)=\Bar{d_A}-\Bar{d_B}
    \quad \text{if} \quad l_A(x)\ge l_B(x)\quad\forall x\in[0,\omega].
\end{equation}

\newpage

\subsection{Graphical illustration}
\begin{figure}[H]
\begin{flushleft} 
\caption{The $W_1$ Wasserstein distance equals the difference in life expectancy at birth when the survivorship functions do not crossover, USA in 2019 compared with USA in 1990, both sexes combined} 
\label{fig_USA}
\vspace{0em} 
\includegraphics[width=\textwidth]{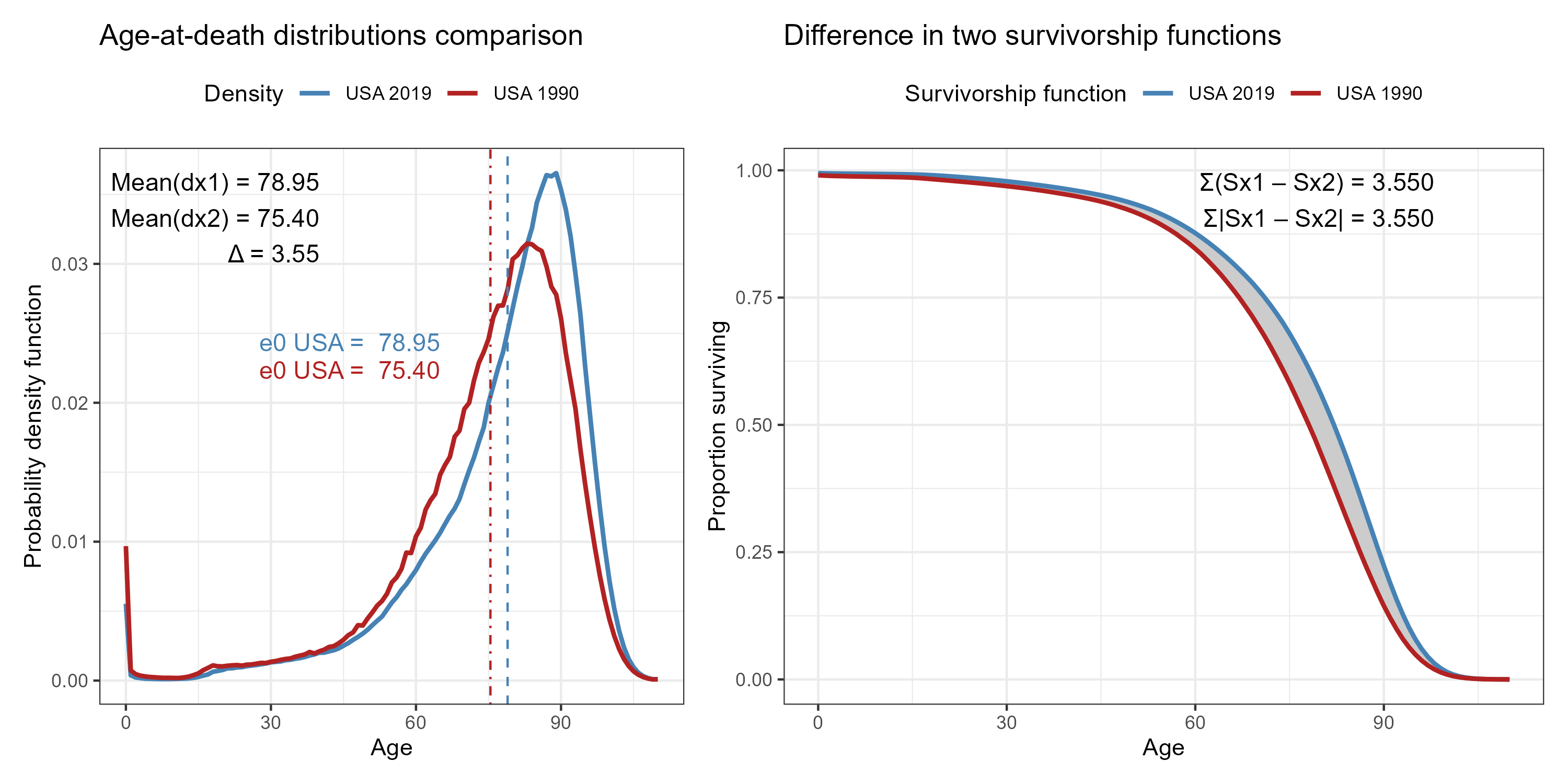} 
\end{flushleft}
\end{figure}

\begin{figure}[H]
\begin{flushleft} 
\caption{The $W_1$ Wasserstein distance differs from the difference in life expectancy at birth when the survivorship functions crossover, USA in 2019 compared with Germany in 1990, both sexes combined} 
\label{fig_Germany}
\vspace{0em} 
\includegraphics[width=\textwidth]{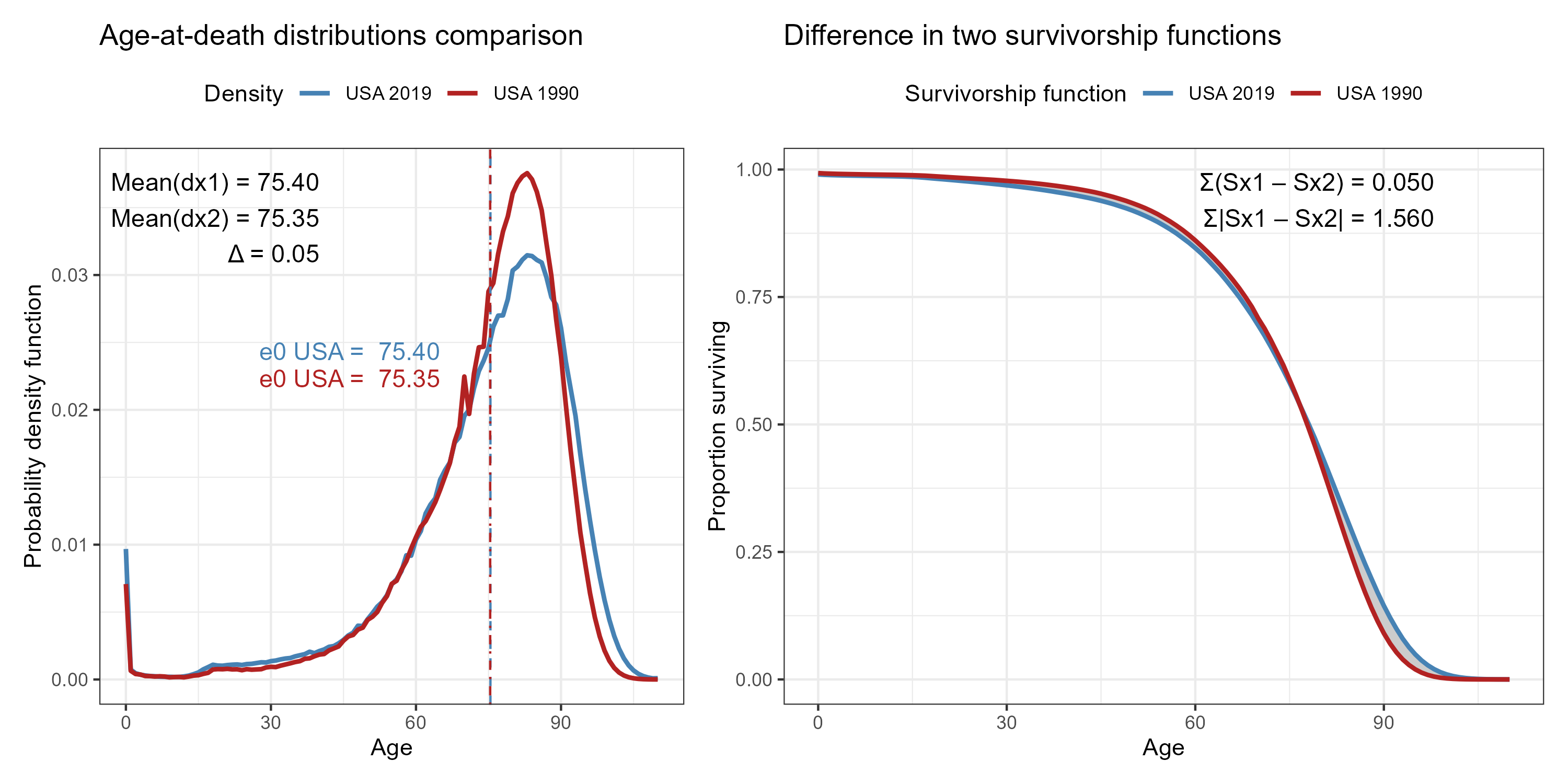} 

\end{flushleft}
\end{figure}

The figures \ref{fig_USA} and \ref{fig_Germany} demonstrate the relationship between the $W_1$ Wasserstein distance and the difference in $e_0$ based on empirical data obtained from the Human Mortality Database \citep{HMD2025}. The first figure depicts a mortality comparison over time (US in 1990 vs. US in 2019). The age-at-death distribution has shifted to older ages as shown in the left panel on figure \ref{fig_USA}. The mean age at death or $e_0$ value increased by 3.55 years (from 75.4 to 78.95 years). The corresponding survivorship function are drawn in the right panel of the figure. There are not crossing over as the survivorship function referring to the year 2019 shows higher proportion of survivors at each age. Therefore, the difference between the two survivorship functions (gray-shaded area) summed over age gives both, the $W_1$ Wasserstein distance and the $e_0$ difference.

Figure \ref{fig_Germany} demonstrates a case where both measures lead to different results. While the US and Germany show a very similar mean ages at death in 1990 (75.40 vs. 75.35 years), the age-at-distributions do not look alike. The distribution for the US is characterized by a steeper shape. Further, the density of deaths at very high ages is slightly larger in the German life table. Accordingly, the corresponding survivorship functions indicate higher proportions of survivors at younger ages for Americans but lower proportions of survivors for the very old. In this situation, the $l(x)$ functions do cross and both measures provide different values. The distributional differences are reflected in the $W_1$ Wasserstein distance. The difference in $e_0$, however, reflects the difference in net survival, i.e., the gap in mean longevity.

\subsection{Using the Wasserstein distance for cause-specific mortality comparisons}
As suggested by \cite{Cilek2023}, the optimal transport framework can be used to study differences in cause-specific mortality distributions. Multi-decrement life tables allow the derivation of age-at-death distributions by specific causes of death \citep{Preston2001}. As before, we set the life table radix $ l_0=1$. Summing $d(x,c)$ over all ages $x$ and causes $c$ therefore equals one, ensuring that the resulting cause-specific age-at-death distribution is a proper probability distribution. The Wasserstein distance for the one-dimensional all-cause distribution can be computed analytically using equation \ref{eq7}. In contrast, the two-dimensional Wasserstein distance does not have a closed-form solution. To obtain the optimal transport plan, we can use the Python Optimal Transport (POT) package, which implements the network simplex algorithm to solve the transport problem.

The two-dimensional Wasserstein distance requires specifying a cost matrix that determines how costly it is to move probability mass between age-cause combinations. As described above, we set the cost of moving mass across ages as the absolute age difference. Choosing the cost of movement between causes of death, however, is less straightforward. Setting the cross-cause cost to zero makes movements between causes free, i.e., the resulting Wasserstein distance would then ignore cause-specific differences and measure only the mismatch in age patterns between two distributions. At the opposite extreme, setting the cost to infinity makes movements between causes impossible. In this case, the algorithm solves the optimal transport problem for each cause separately. Intermediate choices are also possible. For example, setting the cause-movement cost equal to one implies that shifting mass by one year of age has the same cost as reallocating mass across causes. The appropriate scale therefore depends on the substantive interpretation of cause-of-death differences. The choice of the cost for moving mass determines how large the Wasserstein distance can be. Mathematically, we can define the cost matrix for the two-dimensional Wasserstein distance as,

\begin{equation}
M_{ij}=|x_i-x_j\ |+\lambda\mathbf{1}\{c_i\neq c_j\ \},
\end{equation}

where $|x_i-x_j|$ is the transport cost for moving mass between ages (measured in years), while $\lambda\mathbf{1}\{c_i\neq c_j\}$ is an indicator equal to one if the causes differ.
The two-dimensional Wasserstein distance does not have a fixed maximum value because it depends on the size of $\lambda$. A pragmatic strategy is to anchor the cost of moving mass between causes to the all-cause (one-dimensional) Wasserstein distance. This can be done dynamically, i.e., assigning each pair of distributions its own cause-movement cost based on their one-dimensional Wasserstein distance, or statically, using a fixed all-cause Wasserstein distance value for all two-dimensional calculations. If we do not set the cost for moving mass between causes of death to infinity, the two-dimensional Wasserstein distance quantifies how much work is needed to align both age patterns and cause-of-death patterns simultaneously, where work is defined by the amount of death probability mass that must be shifted multiplied by the distance (or cost) over which it is moved.

\begin{figure}[H]
\centering
\caption{Toy example for demonstrating the idea behind theWasserstein distance using cause-specific mortality}
\includegraphics[width=\textwidth,]{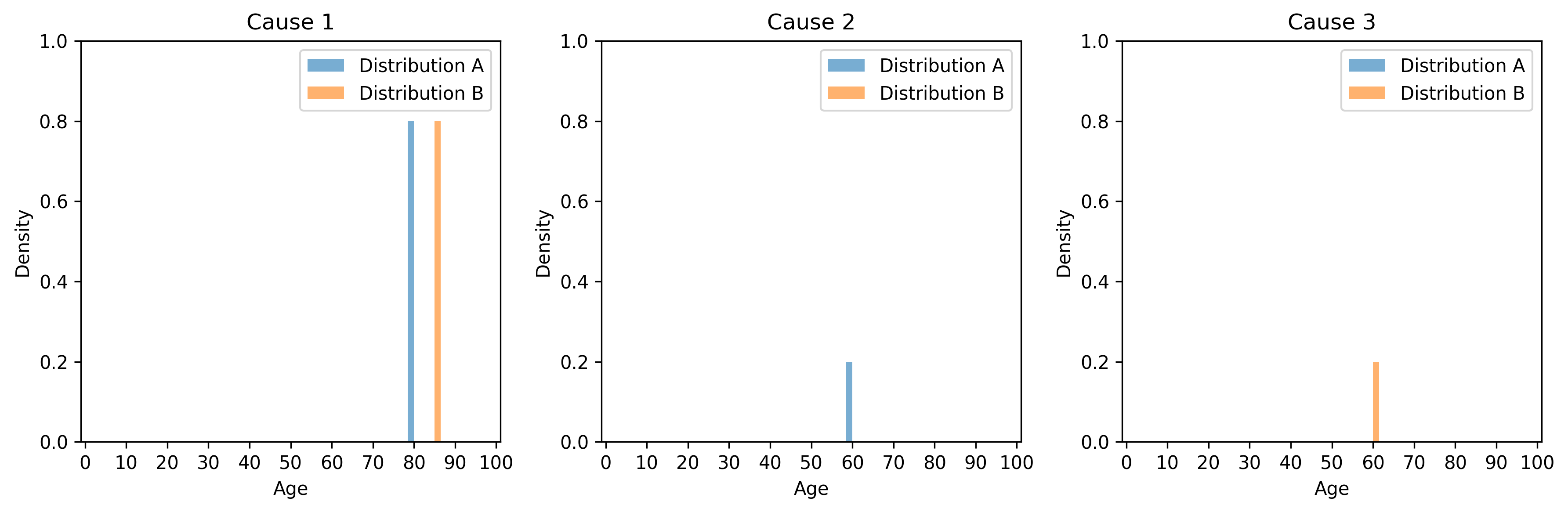}
\label{fig2d}
\medskip
\small
\begin{minipage}{\linewidth}
\raggedright
To make both distributions look the same, we first need to move $0.8$ mass from age $85$ to $80$ $(|80-85|\cdot0.8=4)$. Then, we need to move $0.2$ mass between causes two and three (the age dimension is already equal for both distributions). The key question is how much does moving mass between two causes cost? If we use a $\lambda$ of $10$, the transport is quantified by $|60-60|+10\cdot0.2=2$. Adding both transports together gives a two-dimensional Wasserstein distance value of six $(4+2=6)$.
\end{minipage}
\vspace{-10pt}
\end{figure}

\subsection{Comparing the $W_1$ Wasserstein distance to the non-overlap index}

Several approaches exist for quantifying differences between age-at-death 
distributions. For instance, \cite{Shi2022} introduced the non-overlap index 
(also referred to as the Jaccard similarity index) to measure differences 
between age-at-death distributions. The measure is formally defined as,

\begin{equation}
    J(p, q) = \frac{\int_{-\infty}^{\infty} \min(p(x), q(x))\, dx}
                   {\int_{-\infty}^{\infty} \max(p(x), q(x))\, dx},
\end{equation}

where $p(x)$ and $q(x)$ are two probability density functions.  The index equals one when the two distributions are identical and zero when 
their supports are entirely disjoint, so that larger values indicate greater  similarity. The corresponding Jaccard distance is $JD = 1 - J(p, q)$.

As its name suggests, the non-overlap index quantifies distributional 
differences in terms of the shared mass between two probability densities. 
Since the $W_1$ Wasserstein distance reflects the absolute difference between the corresponding cumulative 
distribution functions both measure might respond similarly to a shift of probability mass toward older ages and thus, track each other closely in demographic 
applications.

To examine this relationship formally, it is convenient to express $J$ through 
the Total Variation (TV) distance. As shown by \cite{Moulton2018}, $J$ can be defined as, 

\begin{equation}\label{eq:JTV}
    J = \frac{1 - \mathrm{TV}}{1 + \mathrm{TV}},
\end{equation}
with
\begin{equation}\label{eq:TV}
    \mathrm{TV}(p, q) 
    = \frac{1}{2} \int_{-\infty}^{\infty} \lvert p(x) - q(x) \rvert\, dx.
\end{equation}

The equation \ref{eq:JTV} shows that $J$ and $\mathrm{TV}$ carry 
identical information. To connect $\mathrm{TV}$ to $W_1$, we consider the demographically relevant case in which $q(x) = p(x - \delta)$ is a pure location shift of $p$ by 
$\delta > 0$ years along the age axis, while the shape of the distribution 
remains unchanged. The TV distance becomes then,

\begin{equation}\label{eq:TVshift}
    \mathrm{TV}(\delta) 
    = \frac{1}{2} \int_{-\infty}^{\infty} \lvert p(x) - p(x - \delta) \rvert\, dx.
\end{equation}

Applying a first-order Taylor expansion, $p(x - \delta) \approx p(x) - \delta\, p'(x)$, gives,

\begin{equation}
    \lvert p(x) - p(x - \delta) \rvert \approx \delta \lvert p'(x) \rvert,
\end{equation}

\noindent so that

\begin{equation}\label{eq:TVapprox}
    \mathrm{TV}(\delta)
    \approx \frac{\delta}{2} \int_{-\infty}^{\infty} \lvert p'(x) \rvert\, dx.
\end{equation}

\noindent For any unimodal density, $p'(x)$ changes sign exactly once, at
the modal age $M$. Hence, we have a rising part, where $p'(x) > 0$,

\begin{equation}
    \int_{-\infty}^{M} p'(x)\, dx = p(M) - p(-\infty) = p(M) - 0 = p(M),
\end{equation}

\noindent and a falling part, where $p'(x) < 0$,

\begin{equation}
    \int_{M}^{\infty} -p'(x)\, dx = -(p(\infty) - p(M)) = -(0 - p(M)) = p(M).
\end{equation}

\noindent The total variation of $p'(x)$ therefore equals twice the modal
density: $\int_{-\infty}^{\infty} \lvert p'(x) \rvert\, dx = 2\,p(M)$.
Substituting into \eqref{eq:TVapprox} yields

\begin{equation}\label{eq:TVlinear}
    \mathrm{TV}(\delta) \approx \delta \cdot p(M).
\end{equation}

Under a pure location shift, it follows directly that, 
\begin{equation}\label{eq:W1delta}
    W_1(p, q)  = \delta.
\end{equation}

Combining \eqref{eq:TVlinear} and \eqref{eq:W1delta} with the
exact identity \eqref{eq:JTV} gives the following approximation for the
Jaccard index in terms of $W_1$,

\begin{equation}\label{eq:Japprox}
    J \approx \frac{1 - W_1 \cdot p(M)}{1 + W_1 \cdot p(M)}.
\end{equation}

Equation~\eqref{eq:Japprox} implies that $J$ is a function of $W_1$ alone, with the modal density $p(M)$
acting as a scaling factor that depends on the shape of the $d(x)$
distribution.

\begin{figure}[H]
\centering
\caption{Comparing the exact Jaccard index with the approximated Jaccard index}
\includegraphics[width=\textwidth,]{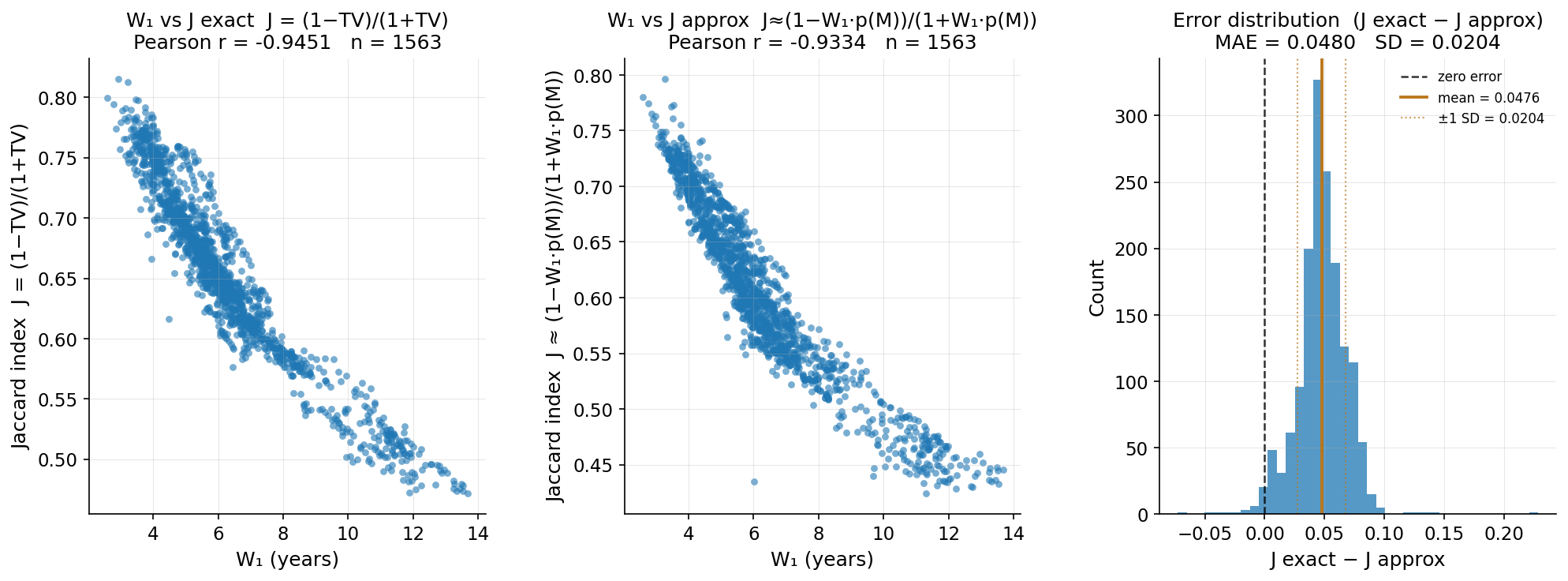}
\label{figJaccardApprox}
\raggedright
The plot compares the exact Jaccard index with the approximated version on the basis of Human Mortality Database life tables for women and men in the period $1990-2024$. The analysis includes 1563 pairs (combinations of year, sex, and country).
The correlation is not linear. Even if $p(M)$ were identical across all pairs, the relationship between $J_{\text{approx}}$ and $W_1$ would not be linear. It is a strictly convex decreasing curve that flattens as $W_1$ grows because the denominator $(1 + W_1 \cdot p(M))$ increases with $W_1$, compressing Jaccard values toward zero for large shifts. 
\vspace{-10pt}
\end{figure}

This chain of relationships, $W_1 = \delta$,
$\mathrm{TV} \approx \delta \cdot p(M)$ from the Taylor expansion, and
$J = (1 - \mathrm{TV})/(1 + \mathrm{TV})$ exactly, suggests that both
measures closely track each other in demographic applications. The
correlation is high whenever mortality improvements are predominantly
location shifts of the $d(x)$ distribution, i.e., deaths are shifted
toward older ages with relatively modest changes in the shape of the
distribution. When shape changes such as the compression of mortality
(rectangularization) are also present, the modal density $p(M)$ varies
across comparisons, introducing an bias to the otherwise
tight $W_1$ to $J$ relationship.

Additionally, distributional differences are often measured by the
Kullback-Leibler (KL) divergence, defined as

\begin{equation}
    \mathrm{KL}(p, q)
    = \int_{-\infty}^{\infty} p(x) \log \frac{p(x)}{q(x)}\, dx.
\end{equation}

The KL divergence quantifies how much one probability distribution
diverges from a reference distribution: it equals zero if and only if $p$
and $q$ are identical, and increases as the two distributions become more
dissimilar. Unlike the non-overlap index, the KL divergence has no direct interpretation in terms of the geometry
of the distributions. Rather than measuring how far apart the two
distributions sit on the age axis, it measures the expected information
gain when distinguishing $p$ from the reference $q$, and is therefore
sensitive to differences in distributional shape.
\section{Decomposing the $W_1$ Wasserstein Distance into shift and dispersion contributions}

Recently, \citet{Resin2025} introduced a quantile-based decomposition of the $W_1$ Wasserstein distance which expresses the measure as four additive terms: a positive and a negative shifting term, and a positive and a negative dispersion term,
\begin{equation}\label{eq:decompo}
    W_1(P,Q) = \mathrm{Shift}^{W_1}_+(P,Q) + \mathrm{Shift}^{W_1}_-(P,Q) + \mathrm{Disp}^{W_1}_+(P,Q) + \mathrm{Disp}^{W_1}_-(P,Q),
\end{equation}
where the positive shifting term captures an upward shift of $P$ relative to $Q$, the negative shifting term reflects a downward shift of $P$ relative to $Q$, the positive dispersion term quantifies the extent to which $P$ is more dispersed than $Q$, and the negative dispersion term reflects the extent to which $P$ is less dispersed than $Q$.

The decomposition is derived by expressing $W_1$ in terms of quantile functions,
\begin{equation}\label{eq:AVM}
    W_1(P,Q) = \int_0^1 \left| F_P^{-1}(u) - F_Q^{-1}(u) \right| du = \frac{1}{2} \int_0^1 \mathrm{AVM}_{\alpha}(P,Q)\, d\alpha,
\end{equation}
with
\begin{equation}
    \mathrm{AVM}_{\alpha}(P,Q) = \left|F_P^{-1}\!\left(\tfrac{1+\alpha}{2}\right) - F_Q^{-1}\!\left(\tfrac{1+\alpha}{2}\right)\right| + \left|F_P^{-1}\!\left(\tfrac{1-\alpha}{2}\right) - F_Q^{-1}\!\left(\tfrac{1-\alpha}{2}\right)\right|.
\end{equation}
\citet{Resin2025} refer to this representation of $W_1$ as the \textit{area validation metric} (AVM). The integrand $\mathrm{AVM}_{\alpha}(P,Q)$ compares the quantile spreads of $P$ and $Q$ at levels $\frac{1-\alpha}{2}$ and $\frac{1+\alpha}{2}$ symmetrically around the median.

The pointwise decomposition terms are defined as
\begin{align}
    \mathrm{Shift}^{W_1}_{\alpha,+}(P,Q) &:= 2\left[\min\!\left(F_P^{-1}\!\left(\tfrac{1+\alpha}{2}\right) - F_Q^{-1}\!\left(\tfrac{1+\alpha}{2}\right),\; F_P^{-1}\!\left(\tfrac{1-\alpha}{2}\right) - F_Q^{-1}\!\left(\tfrac{1-\alpha}{2}\right)\right)\right]_+, \\[6pt]
    \mathrm{Disp}^{W_1}_{\alpha,+}(P,Q) &:= \left[\left(F_P^{-1}\!\left(\tfrac{1+\alpha}{2}\right) - F_Q^{-1}\!\left(\tfrac{1+\alpha}{2}\right)\right) - \left(F_P^{-1}\!\left(\tfrac{1-\alpha}{2}\right) - F_Q^{-1}\!\left(\tfrac{1-\alpha}{2}\right)\right)\right]_+,
\end{align}
The two negative components are defined symmetrically by swapping $P$ and $Q$,
\begin{equation}
    \mathrm{Shift}^{W_1}_-(P,Q) := \mathrm{Shift}^{W_1}_+(Q,P) \qquad \text{and} \qquad \mathrm{Disp}^{W_1}_-(P,Q) := \mathrm{Disp}^{W_1}_+(Q,P).
\end{equation}
Integrating the pointwise terms as in \eqref{eq:AVM} yields the final decomposition components,
\begin{equation}
    \mathrm{Shift}^{W_1}_{\pm}(P,Q) := \frac{1}{2}\int_0^1 \mathrm{Shift}^{W_1}_{\alpha,\pm}(P,Q)\, d\alpha \qquad \text{and} \qquad \mathrm{Disp}^{W_1}_{\pm}(P,Q) := \frac{1}{2}\int_0^1 \mathrm{Disp}^{W_1}_{\alpha,\pm}(P,Q)\, d\alpha.
\end{equation}

Table~\ref{tab:wd_decomp} illustrates the decomposition for two contrasting examples. Applying the decomposition to age-at-death distributions for Germany in 1990 and 2019 suggests that the two distributions differ predominantly through a shift of deaths towards older ages, with the shift contribution ($4.64$ years) accounting for the majority of the total $W_1$ distance, and a comparatively modest dispersion contribution ($1.20$ years). Notably, the $W_1$ distance of $5.85$ years coincides exactly with the difference in life expectancy at birth between the two years, indicating that the corresponding $lx$ functions do not cross. The small residual dispersion component reflects a modest compression of mortality in 2019 relative to 1990.

In contrast, England \& Wales and Iceland in 1849 share an identical life expectancy ($e_0 = 37.34$ years). Yet their age-at-death distributions differ substantially, yielding a $W_1$ distance of $7.33$ years driven almost entirely by dispersion ($7.33$ years) with a negligible shift ($0.01$ years). In this example the corresponding $lx$ functions crossover and hence, the difference in $e_0$ does not reflect the large distributional differences correctly.

\begin{table}[ht]
    \centering
    \caption{Wasserstein distance decomposition for two contrasting mortality comparisons. Shift and dispersion contributions sum to $W_1$. All values are in years.}
    \label{tab:wd_decomp}
    \begin{tabular}{lcccccc}
        \toprule
        Comparison & $e_0^A$  & $e_0^B$  & $\Delta e_0$ & $W_1$ & Shift & Dispersion \\
        \midrule
        Germany 1990 vs.\ 2019           & 75.35 & 81.20 & 5.85 & 5.85 & 4.64 & 1.20 \\
        England \& Wales vs.\ Iceland, 1849 & 37.34 & 37.34 & 0.00 & 7.33 & 0.01 & 7.33 \\
        \bottomrule
    \end{tabular}
\end{table}

\section{Results}
The following analysis uses period life tables from the \href{https://www.mortality.org/}{Human Mortality Database}. The data can be downloaded for free after creating an account. The Python code used for conducting the following analysis is available at \href{https://github.com/msauerberg/Wasserstein_repo}{github.com/msauerberg}.

\subsection{Sampling from the Human Mortality Database}
\begin{figure}[H]
    \centering
    \caption{The relationship between the $W_1$ Wasserstein distance and differences in life expectancy at birth}
    \begin{subfigure}[b]{0.45\textwidth}
        \centering
        \caption{Distribution of the $W_1$ Wasserstein distance and differences in life expectancy at birth for 5 000 samples}
        \includegraphics[width=\textwidth, height=6cm]{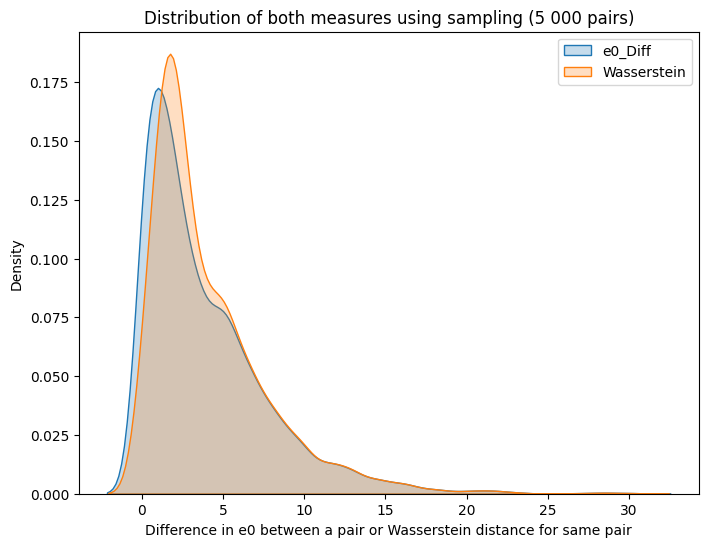}
        \label{fig1a}
    \end{subfigure}
    \hfill
    \begin{subfigure}[b]{0.45\textwidth}
        \centering
        \caption{Scatterplot of the $W_1$ Wasserstein distance vs. differences in life expectancy at birth}
        \includegraphics[width=\textwidth, height=6cm]{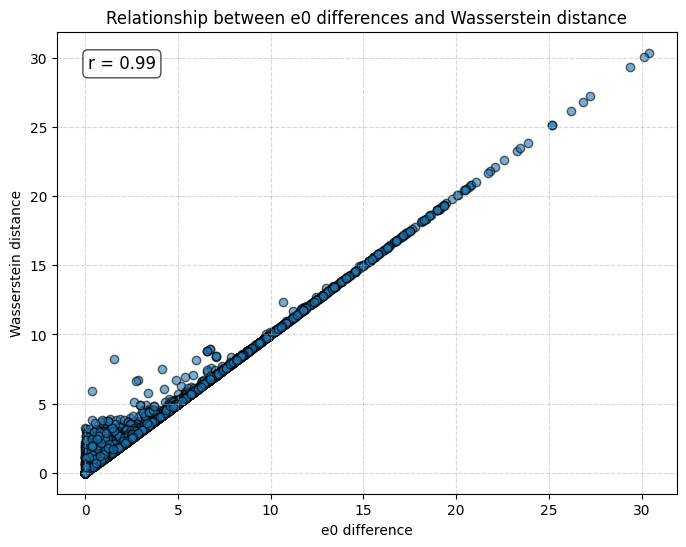}
        \label{fig1b}
    \end{subfigure}
    \label{fig1}
    \vspace{-10pt}
\end{figure}

\begin{figure}[H]
\centering
\caption{Comparing the $W_1$ Wasserstein distance with the Kullback-Leibler divergence and the non-overlap index (Jaccard distance)}
\includegraphics[width=\textwidth,]{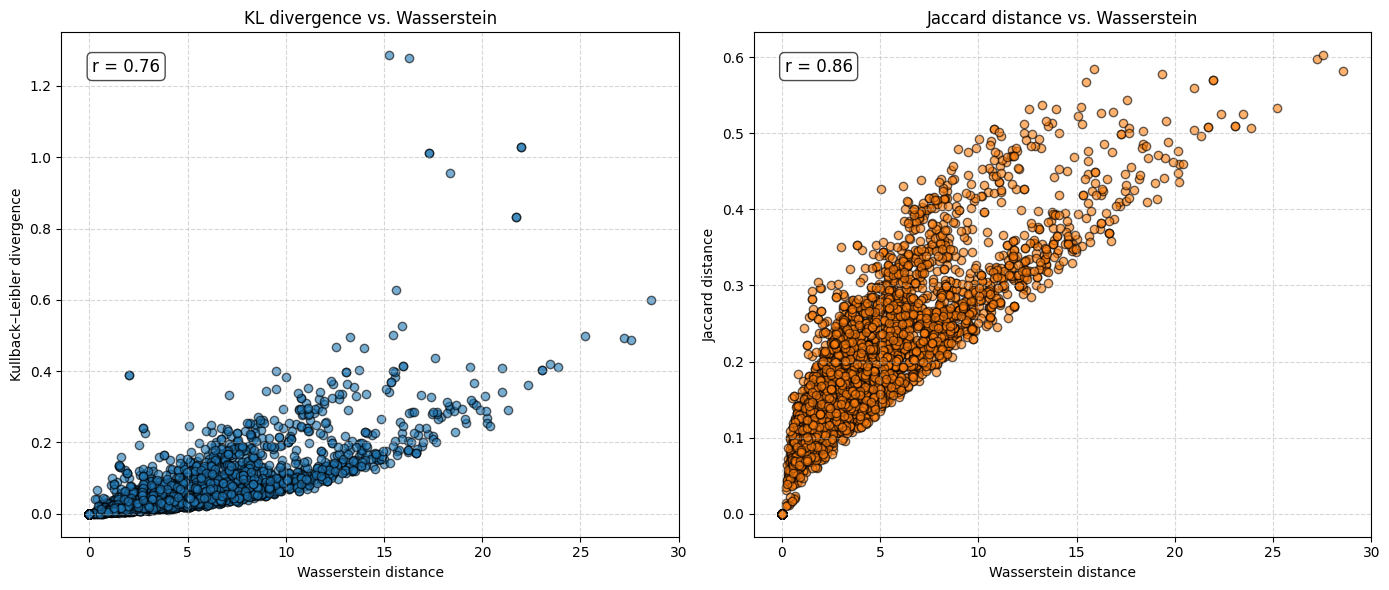}
\label{fig2}
\vspace{-10pt}
\end{figure}

The database includes period life tables for several countries in various years (sometimes going back to the 1800s). Each period life table is available for women, men, and for both sexes combined. It is computationally demanding to compare differences in $e_0$ and $W_1$ Wasserstein distance for all possible pairs in the dataset. For this reason, we focus on life tables for both sexes combined and sample 5 000 randomly selected country pairs from the subset. The cross-country comparison refers to one randomly select year, i.e., we do not compare period life tables over time. In addition to $W_1$ and differences in $e_0$, we calculate the non-overlap index - also called Jaccard distance - \citep{Shi2022} and the Kullback-Leibler divergence for each pair. This allows comparing the $W_1$ Wasserstein result to two alternatives measures for distributional similarity. One of the 5 000 sampled pairs is Denmark and Belgium in 1887. For this pair, the difference in $e_0$ equals the $W_1$ Wasserstein distance ($3.33$). The non-overlap index is $0.14$, while the Kullback-Leibler divergence is $0.02$. The results for all 5 000 pairs is depicted in Figure \ref{fig1} and \ref{fig2}.

The distributions of the $W_1$ Wasserstein distance and the difference in $e_0$ largely overlap. Both, the $W_1$ Wasserstein distance and $e_0$ differences range from 0 to 30.38 years (Figure \ref{fig1a}). The mean is slightly larger for $W_1$ as compared to $e_0$ differences (4.18 vs. 3.95). The scatterplot in Figure \ref{fig1b} shows an extremely strong correlation between the two measures (Pearson’s r=0.99). Moreover, $W_1$ also suggests a strong relationship with the two alternative measures of distributional difference (see Figure \ref{fig2}).

\begin{figure}[H]
    \centering
    \caption{Largest gap in life expectancy at birth and largest $W_1$ Wasserstein distance}
    \begin{subfigure}[b]{0.45\textwidth}
        \centering
        \caption{Age-at-death distribution for Denmark and Italy in 1918}
        \includegraphics[width=\textwidth]{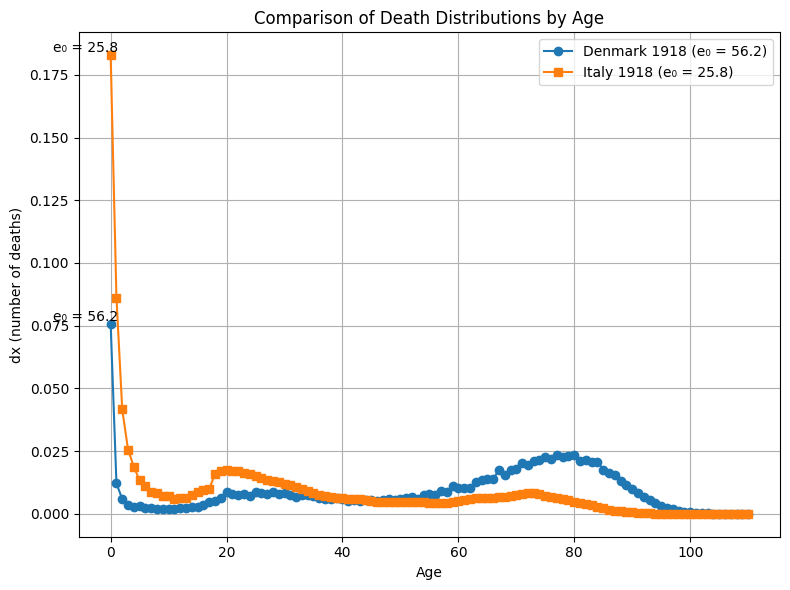}
        \label{fig3a}
    \end{subfigure}
    \hfill
    \begin{subfigure}[b]{0.45\textwidth}
        \centering
        \caption{Survivorship functions for Denmark and Italy in 1918}
        \includegraphics[width=\textwidth]{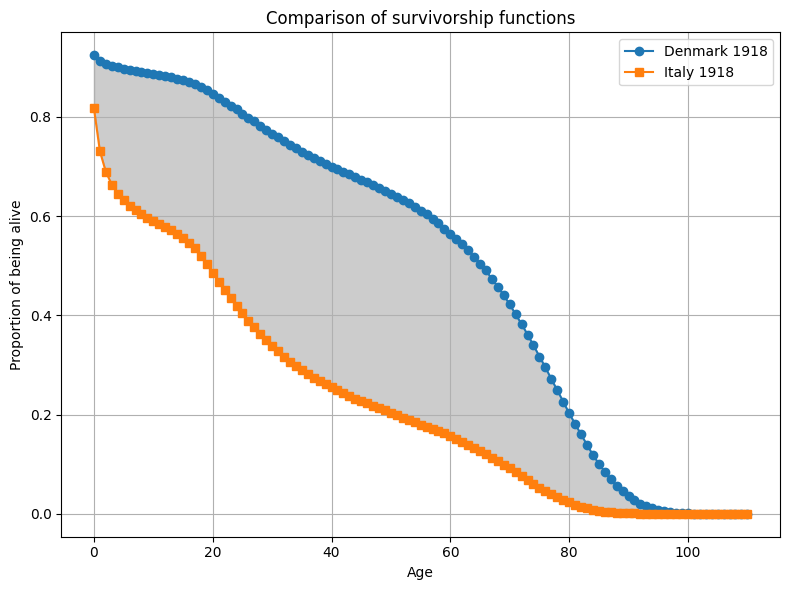}
        \label{fig3b}
    \end{subfigure}
    \label{fig3}
    \vspace{-10pt}
\end{figure}

The largest gap in $e_0$ is observed between Denmark and Italy in 1918, amounting to about 30 years. This pair also shows the highest observed $W_1$ Wasserstein distance (Figure \ref{fig3a}). As shown in Figure \ref{fig3b}, Italy experienced particularly high infant mortality, which generated a large survivorship gap across the entire age span.

\begin{figure}[H]
    \centering
    \caption{Small gap in life expectancy at birth but a large $W_1$ Wasserstein distance}
    \begin{subfigure}[b]{0.45\textwidth}
        \centering
        \caption{Age-at-death distribution for England \& Wales and Iceland in 1849}
        \includegraphics[width=\textwidth]{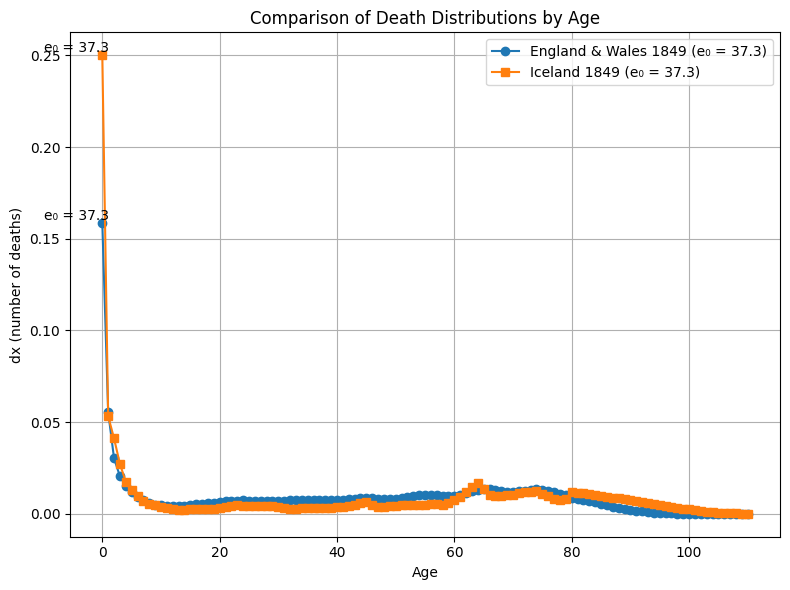}
        \label{fig4a}
    \end{subfigure}
    \hfill
    \begin{subfigure}[b]{0.45\textwidth}
        \centering
        \caption{Survivorship functions for England \& Wales and Iceland in 1849}
        \includegraphics[width=\textwidth]{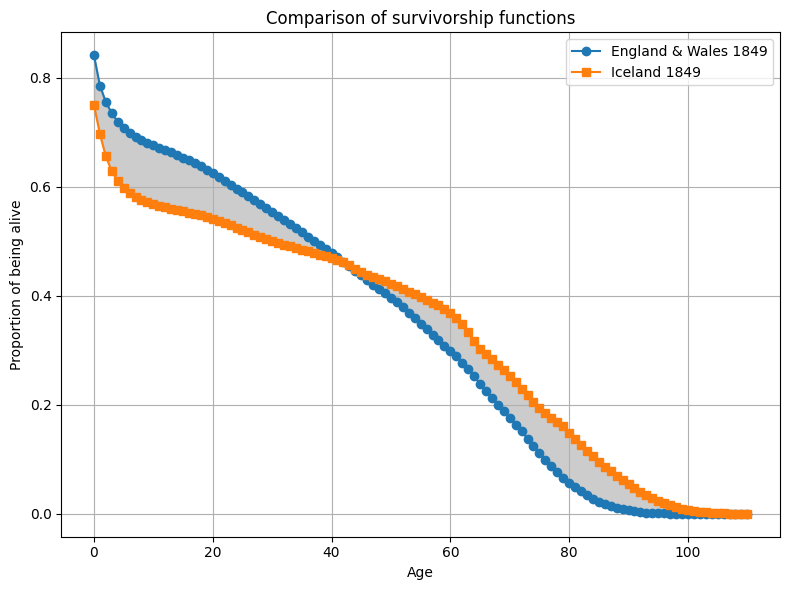}
        \label{fig4b}
    \end{subfigure}
    \label{fig4}
    \vspace{-10pt}
\end{figure}

In most cases, a large $e_0$ difference coincides with a large $W_1$ Wasserstein distances. However, we do find cases where the gap in $e_0$ is very small but $W_1$ Wasserstein suggests large distributional differences. This is observed, for example, when comparing England \& Wales with Iceland in 1849. Both countries show a $e_0$ value of about 37.3. Yet, the age-at-death distributions differ substantially (see Figure \ref{fig4}). While Iceland shows higher infant mortality as compared to England \& Wales, mortality is lower for Iceland at older ages. Accordingly, the net difference in survivorship over age is small but there are still large absolute differences in $l(x)$ (see \ref{fig4a}).

\subsection{Analyzing the time period 1990 to 2020}
\begin{figure}[H]
    \centering
    \caption{Distribution of the $W_1$ Wasserstein distance and differences in life expectancy at birth, 1990 to 2020}
    \begin{subfigure}[b]{0.45\textwidth}
        \centering
        \caption{Women}
        \includegraphics[width=\textwidth]{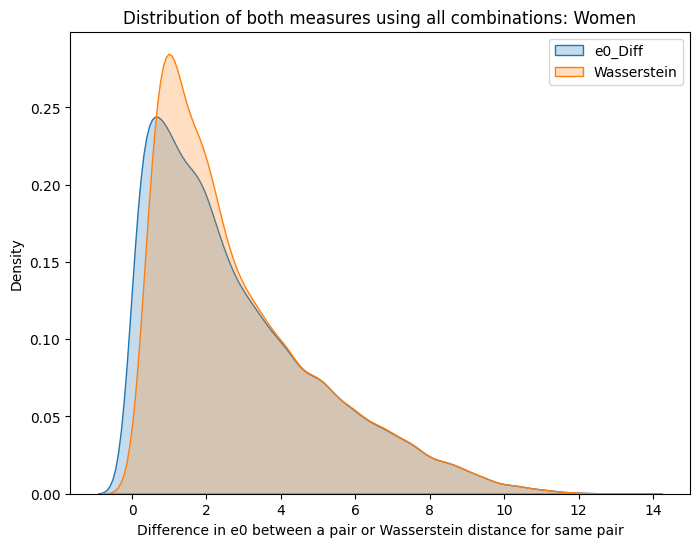}
        \label{fig5a}
    \end{subfigure}
    \hfill
    \begin{subfigure}[b]{0.45\textwidth}
        \centering
        \caption{Men}
        \includegraphics[width=\textwidth]{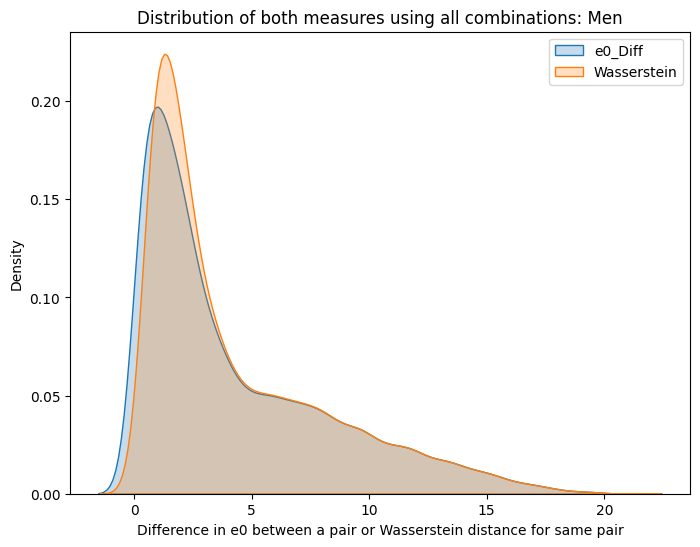}
        \label{fig5b}
    \end{subfigure}
    \label{fig5}
    \vspace{-10pt}
\end{figure}

In more recent periods, the relationship between differences in $e_0$ and the $W_1$ Wasserstein distance is even stronger. Figure \ref{fig5} shows the results for all country pairs from 1990 to 2020, presented separately for women and men. The distributions overlap strongly for both women and men (Figure \ref{fig5a}). The largest $W_1$ Wasserstein distance and $e_0$ differences observed are 13.38 years for women and 20.93 for men (see Table \ref{tab:summary_stats}). The mean of the distribution is slightly larger for $W_1$ as compared to $e_0$ differences (2.92 vs. 2.81 for women and 4.45 vs. 4.33 for men).  

When comparing the difference in $e_0$ with the $W_1$ Wasserstein distance for each pair, the average absolute difference between them is below 0.5 in every year (Figure \ref{fig5b}). The largest observed absolute difference is 3.5, occurring in 1995 for women.  

\begin{table}[H]
\centering
\caption{\vspace{0.3cm}Summary statistics (min, mean, max) for $W_1$ Wasserstein and $e_0$ differences, 1990 to 2020, by sex}
\begin{tabular}{lccc ccc}
\toprule
       & \multicolumn{3}{c}{$W_1$ Wasserstein} & \multicolumn{3}{c}{$e_0$ difference} \\
\cmidrule(lr){2-4} \cmidrule(lr){5-7}
Sex    & Min  & Mean & Max  & Min   & Mean & Max  \\
\midrule
Men    & 0.00 & 4.45 & 20.93 & 0.00 & 4.33 & 20.93 \\
Women  & 0.00 & 2.92 & 13.38 & 0.00 & 2.81 & 13.38 \\
\bottomrule
\end{tabular}
\label{tab:summary_stats}
\end{table}

\begin{figure}[H]
    \centering
    \caption{The absolute difference between the $W_1$ Wasserstein distance and differences in life expectancy at birth}
    \begin{subfigure}[b]{0.45\textwidth}
        \centering
        \caption{Women}
        \includegraphics[width=\textwidth]{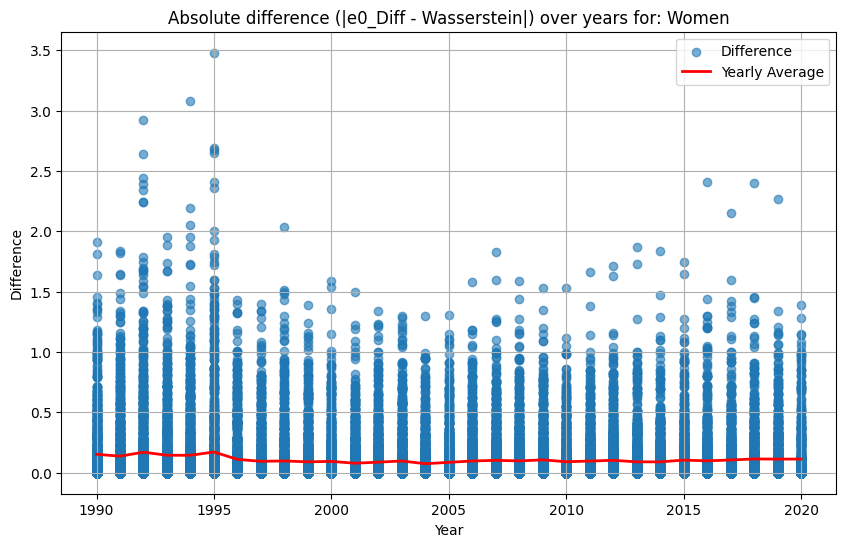}
        \label{fig6a}
    \end{subfigure}
    \hfill
    \begin{subfigure}[b]{0.45\textwidth}
        \centering
        \caption{Men}
        \includegraphics[width=\textwidth]{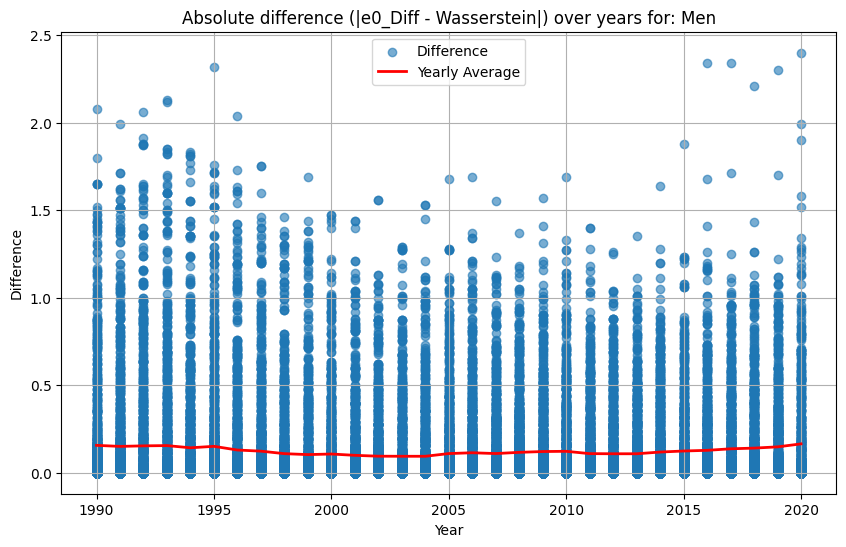}
        \label{fig6b}
    \end{subfigure}
    \label{fig6}
    \vspace{-10pt}
\end{figure}

\subsection{Sex differences in mortality}
\begin{figure}[H]
    \centering
    \caption{The relationship between the $W_1$ Wasserstein distance and differences in life expectancy at birth, sex differences, 1990 to 2020}
    
    \begin{subfigure}[t]{0.45\textwidth}
        \centering
        \caption{Distribution of the $W_1$ Wasserstein distance and differences in life expectancy at birth}
        \includegraphics[width=\textwidth, height=6cm]{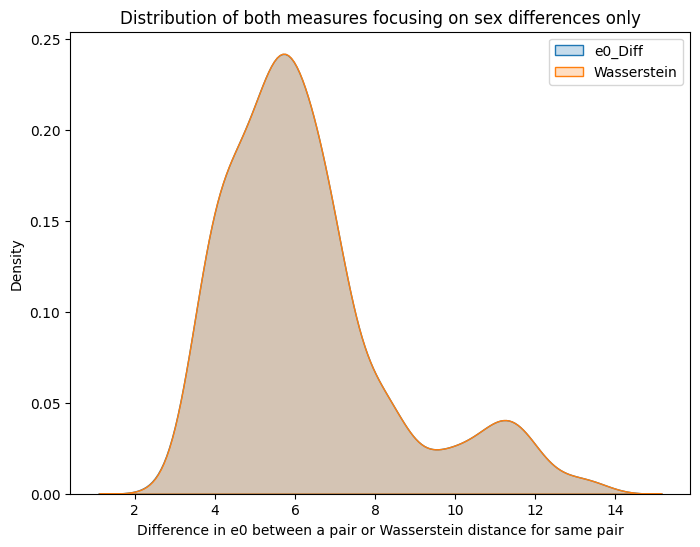}
        \label{fig7a}
    \end{subfigure}
    \hfill
    \begin{subfigure}[t]{0.45\textwidth}
        \centering
        \caption{The absolute difference between the $W_1$ Wasserstein distance and differences in life expectancy at birth}
        \includegraphics[width=\textwidth, height=6cm]{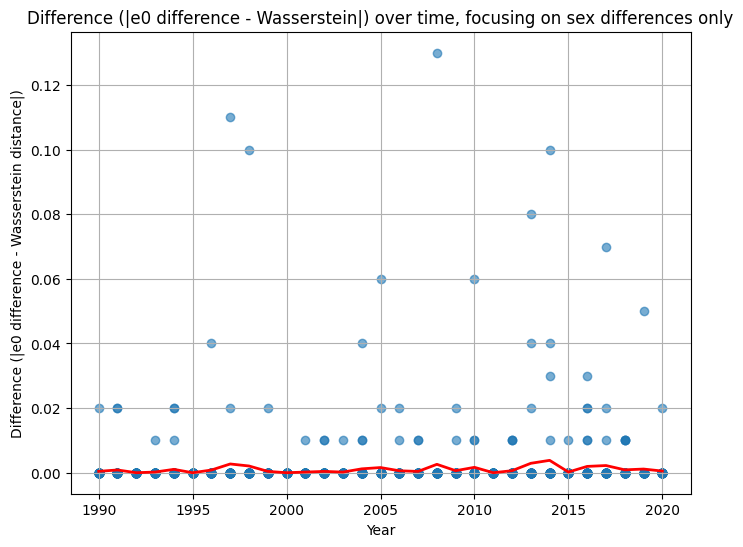}
        \label{fig7b}
    \end{subfigure}
    
    \label{fig7}
\end{figure}

Finally, we compare sex differences in mortality on the basis of $e_0$ differences and the $W_1$ Wasserstein distance. Since women usually show lower death rates at all ages, we can assume that $l_{women}(x) \geq l_{men}(x)$ for all $x$. As described above, the $W_1$ Wasserstein distance equals the difference in $e_0$ under this condition. The presented results based on data for the years 1990 to 2020 confirms that empirically. Figure \ref{fig7a} shows no differences in the distribution for $W_1$ Wasserstein distances and $e_0$ differences. Further, Figure \ref{fig7b} reveals that the absolute difference between both measures is on average smaller than 0.02 in each year. The maximum difference is observed in 2008 with about 0.13 years.

\subsection{Cohort life tables and over time comparisons}
\begin{figure}[H]
    \centering
    \caption{Distribution of the $W_1$ Wasserstein distance and differences in life expectancy at birth for cohorts born between 1890 to 1920}
    \begin{subfigure}[b]{0.45\textwidth}
        \centering
        \caption{Women}
        \includegraphics[width=\textwidth]{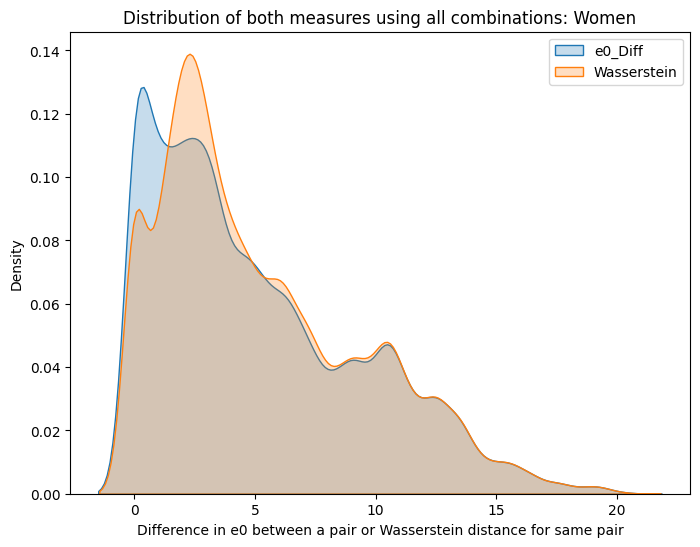}
        \label{fig8a}
    \end{subfigure}
    \hfill
    \begin{subfigure}[b]{0.45\textwidth}
        \centering
        \caption{Men}
        \includegraphics[width=\textwidth]{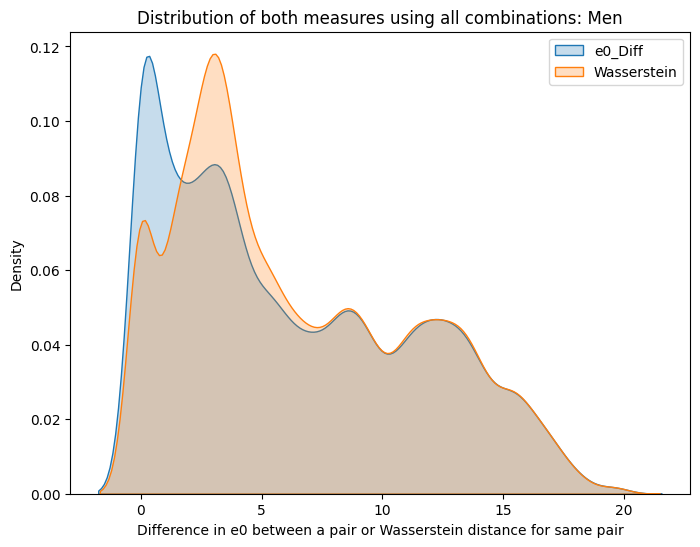}
        \label{fig8b}
    \end{subfigure}
    \label{fig8}
\end{figure}

So far, the analysis used only period life tables and cross-country comparisons in a specific calendar year. To see see whether using cohort life tables and comparisons over time leads to different results, we download the cohort life tables for Denmark, Finland, France, Iceland, Italy, Netherlands, Norway, Spain, Sweden, Switzerland, and England \& Wales from the Human Mortality Database. They describe the mortality experience of all birth cohorts that were born in the 1890s up to the 1920s. We calculate the statistics of interest for 100 730 pairs. The relationship between the $W_1$ Wasserstein distance and differences in $e_0$ measured through Pearson's r is still very strong (r = 0.99 for both, women and men). Also, the distributions are still overlapping but it's less pronounced as compared to the results based on period life tables in specific years (see Figure \ref{fig8}). The summary statistics are shown in Table \ref{tab:summary_stats_cohort}. Again, the mean of the distribution is slightly higher for the $W_1$ Wasserstein distance.  

\begin{figure}[H]
\centering
\caption{Comparing the $W_1$ Wasserstein distance with the Kullback-Leibler divergence and the non-overlap index (Jaccard distance) using cohort data, women}
\includegraphics[width=\textwidth,]{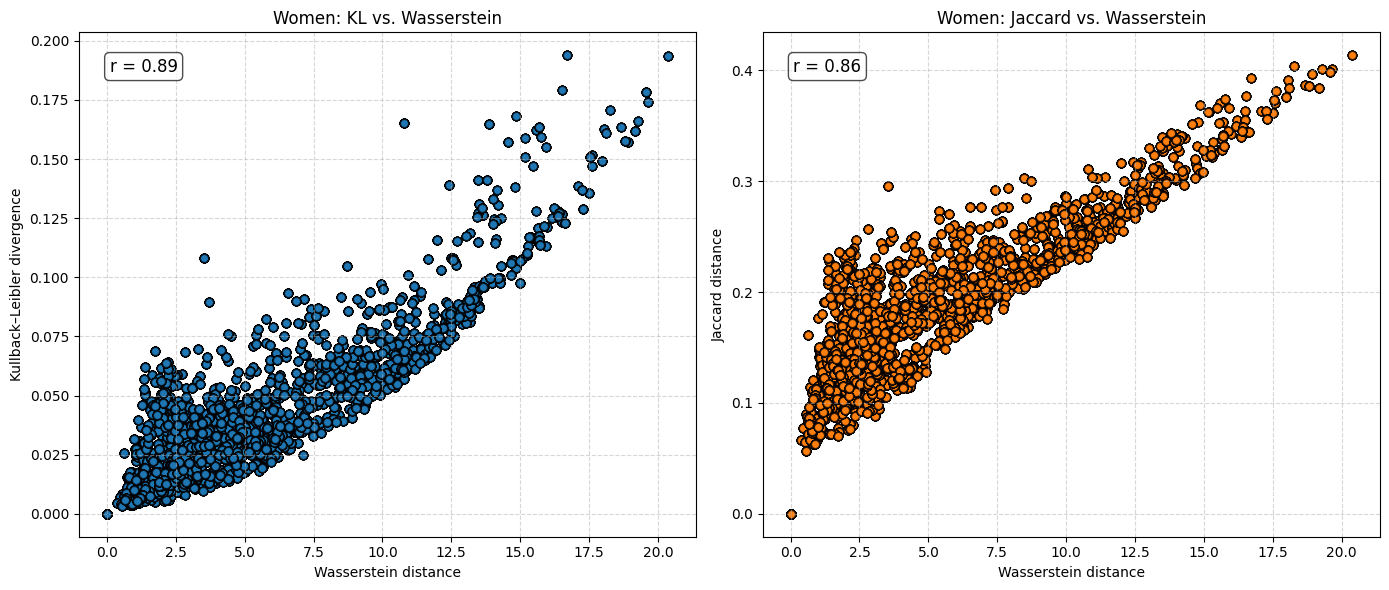}
\label{fig9a}
\end{figure}

\begin{figure}[H]
\centering
\caption{Comparing the $W_1$ Wasserstein distance with the Kullback-Leibler divergence and the non-overlap index (Jaccard distance) using cohort data, men}
\includegraphics[width=\textwidth,]{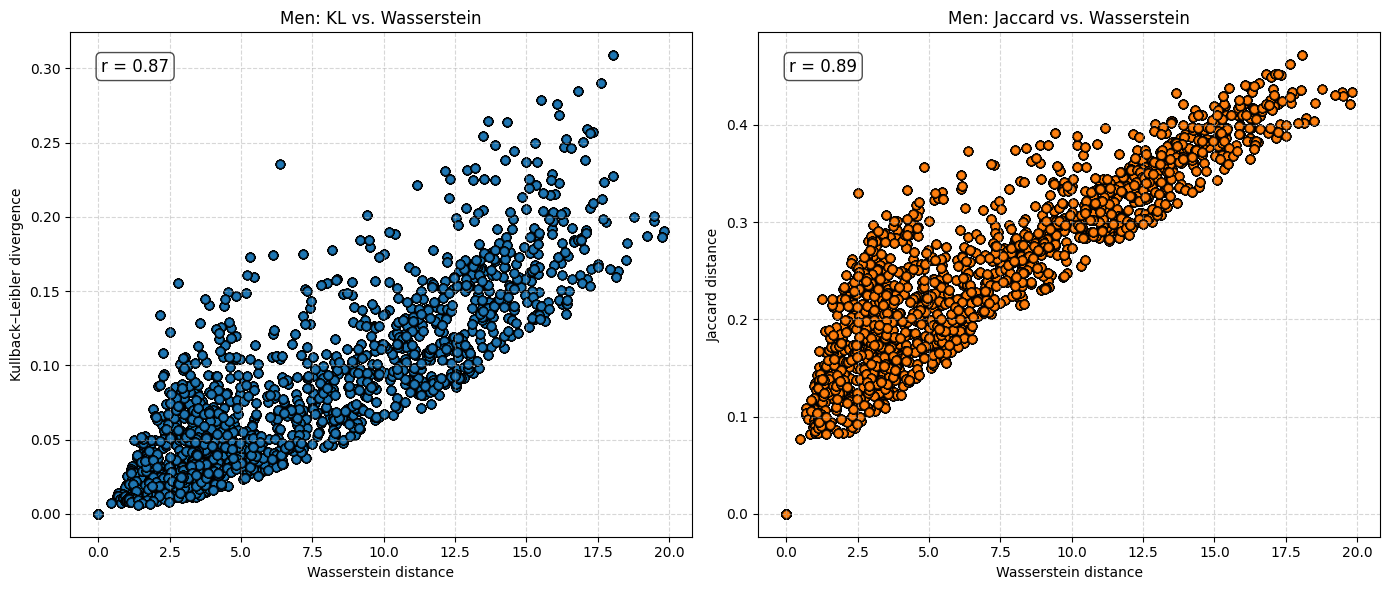}
\label{fig9b}
\end{figure}

This is not surprising because $W_1$ reflects the absolute differences between the $l(x)$ functions. In other words, the $e_0$ difference for a pair of countries cannot be larger than its $W_1$ Wasserstein distance. 
After taking into account comparisons over time, the correlation between the $W_1$ Wasserstein distance and the two alternative measures (KL-divergence and the non-overlap index) remains strong. This is depicted in figure \ref{fig9a} and \ref{fig9b}.

\begin{table}[H]
\centering
\caption{\vspace{0.3cm}Summary statistics (min, mean, max) for $W_1$ Wasserstein and $e_0$ differences for cohorts being born between 1890 and 1920, by sex}
\begin{tabular}{lccc ccc}
\toprule
       & \multicolumn{3}{c}{$W_1$ Wasserstein} & \multicolumn{3}{c}{$e_0$ difference} \\
\cmidrule(lr){2-4} \cmidrule(lr){5-7}
Sex    & Min  & Mean & Max  & Min   & Mean & Max  \\
\midrule
Men    & 0.00 & 6.50 & 19.81 & 0.00 & 6.24 & 19.81 \\
Women  & 0.00 & 5.29 & 20.37 & 0.00 & 5.12 & 20.37 \\
\bottomrule
\end{tabular}
\label{tab:summary_stats_cohort}
\end{table}

\subsection{Comparing cause-specific mortality distributions with the 2-dimensional Wasserstein distance}
\begin{figure}[H]
\centering
\caption{Comparing the $W_1$ Wasserstein distance with the 2-dimensional Wasserstein distance, France and USA}
\includegraphics[width=\textwidth,]{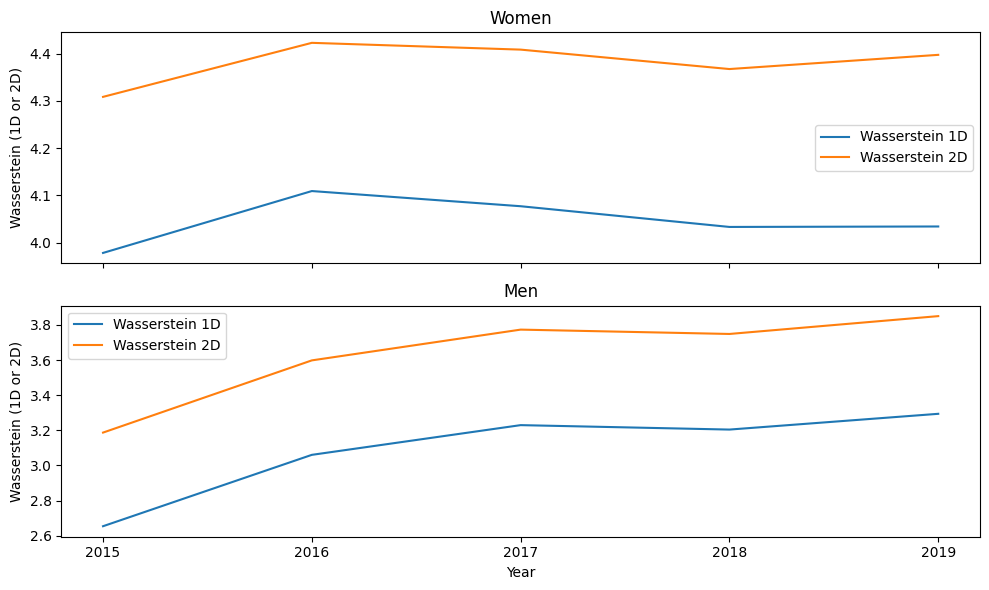}
\label{fig10}
\end{figure}
The figure \ref{fig10} depicts the 1- and 2-dimensional Wasserstein distances for France and the US between 2015 and 2019, separated by women and men. The  $\lambda$ parameter is set to 5, indicating that moving mass between causes of death costs 5 units. The larger  $\lambda$, the larger the 2-dimensional Wasserstein distance and thus, the gap between the 1- and 2-dimensional Wasserstein distances. Interestingly, the 2-dimensional Wasserstein distance is larger for women than for man, while the 1-dimensional case suggests larger differences in the all-cause-specifc age-at-death distributions for men.

\subsection{Decomposing $W_1$ Wasserstein distance into shift- and dispersion contribution}

\begin{figure}[H]
\centering
\caption{Decomposing the $W_1$ Wasserstein distance into shift- and dispersion contribution in three different periods}
\includegraphics[width=\textwidth,]{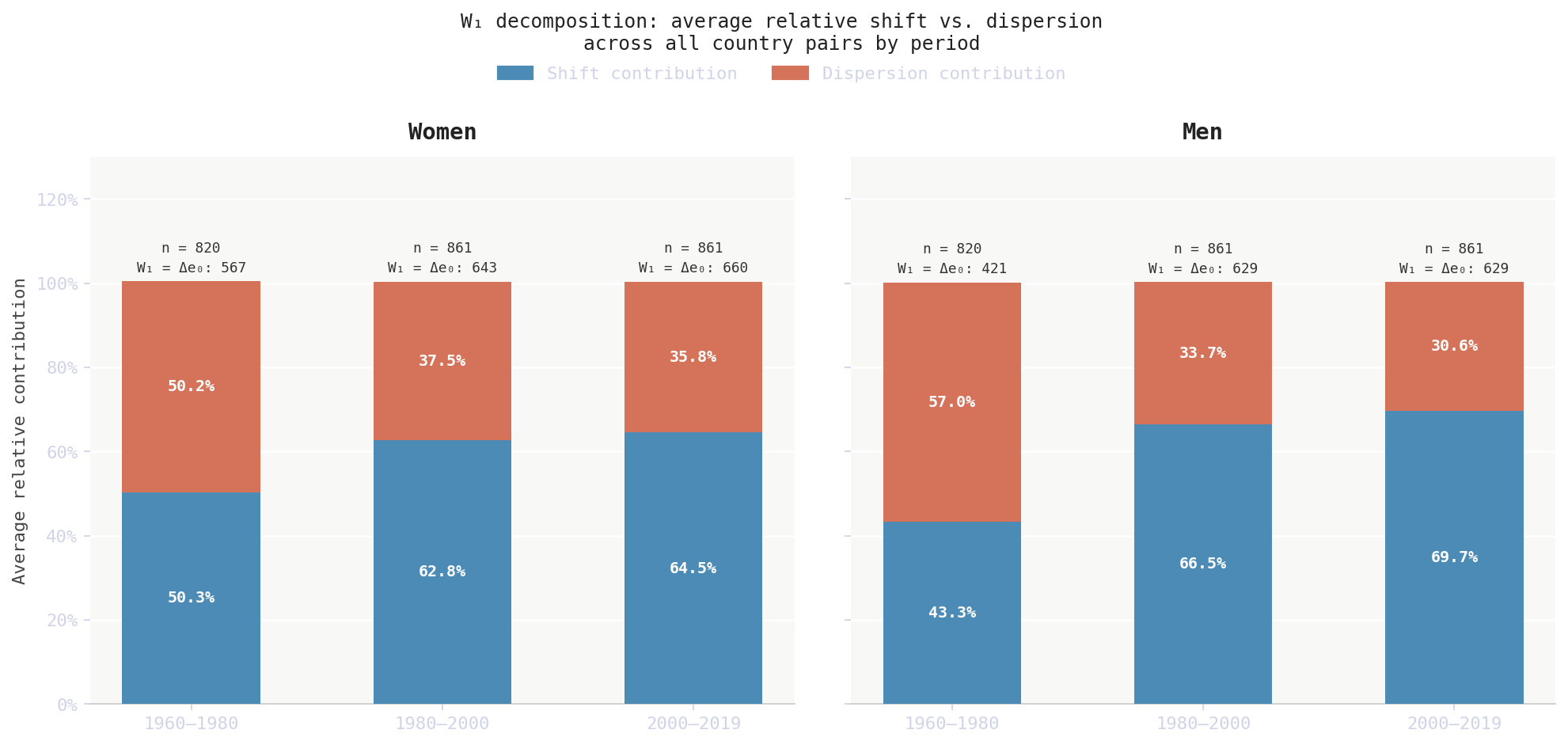}
\label{figdecompo}
\end{figure}

To examine whether differences between age-at-death distributions are driven primarily by a shift of deaths towards older ages or by changes in the shape of the distribution, we apply the decomposition described above (see Equation~\ref{eq:decompo}). Combining the positive and negative shifting and dispersion terms yields a two-component decomposition of the $W_1$ Wasserstein distance into a total shift and a total dispersion contribution.
Figure~\ref{figdecompo} shows the relative contributions averaged across all country pairs with available data in each period. For the period 1960 to1980, roughly half of the average $W_1$ distance is attributable to shifting and half to dispersion among women, while among men the dispersion contribution is slightly larger (57 percent). These averages are based on 861 country pairs. In the majority of cases - 576 among women and 421 among men - the $W_1$ Wasserstein distance equals the difference in life expectancy at birth, $\Delta e_0$. In more recent periods, the shift contribution increases and so does the number of pairs for which $W_1 = \Delta e_0$, reflecting a growing tendency for mortality improvements to take the form of a parallel shift of the age-at-death distribution rather than a change in its spread.

\begin{figure}[H]
\centering
\caption{Decomposing the $W_1$ Wasserstein distance into shift- and dispersion contribution, sex differences}
\includegraphics[width=\textwidth,]{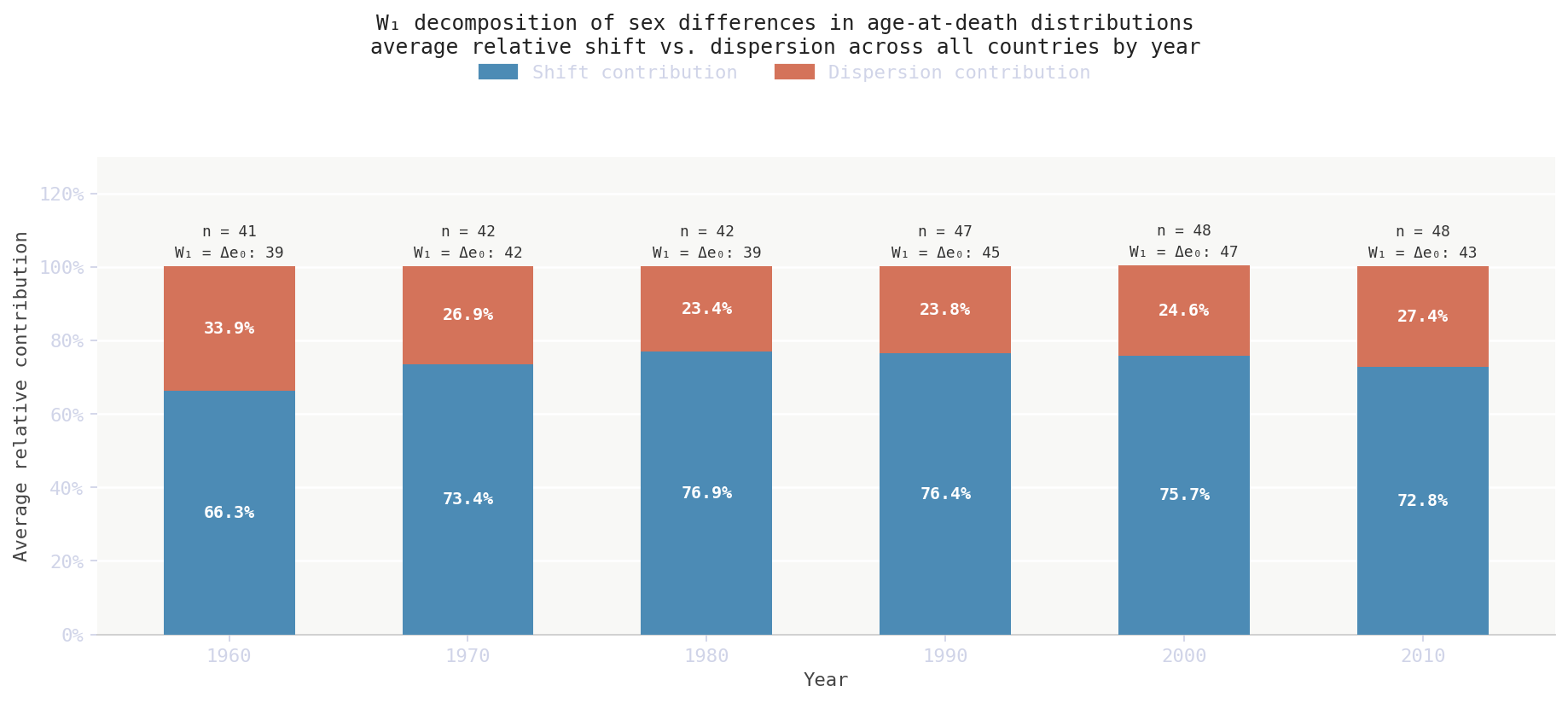}
\label{figdecompo_sex}
\end{figure}

Figure~\ref{figdecompo_sex} shows the decomposition of sex differences in age-at-death distributions across six calendar years. In each year, the $W_1$ Wasserstein distance between the female and male life table distributions is computed for all countries with available data. The results consistently indicate that a shift of deaths towards older ages is the dominant driver of sex differences in $W_1$, with the dispersion contribution remaining small throughout. Moreover, in almost all country-year combinations, $W_1 = \Delta e_0$, meaning that the sex gap in the Wasserstein distance is fully explained by the gap in life expectancy. This identity is assessed by rounding both quantities to one decimal place, rounding to the nearest integer would yield an even higher count of exact agreements.

\section{Conclusion}

When the survivorship functions of two populations do not cross, the $W_1$ Wasserstein distance is equal to the difference in $e_0$. In such cases, the comparison of two age-at-death distributions is no longer simply a comparison of means, but rather the solution to an optimal transport problem. This perspective provides a novel interpretation of differences in life expectancy at birth.
Further, the $W_1$ Wasserstein distance might be used to study distributional differences between $e_0$ and health expectancy at birth. By definition, the survival curve for healthy survivors is smaller or equal to the conventional survival curve. Accordingly, their $W_1$ Wasserstein distance is simply the difference between life expectancy at birth and the health expectancy at birth.

It also possible to use the Wasserstein distance in a two-dimensional setting. This requires the use of an algorithm to solve the optimal transport problem as there is no closed-form solution. Also, it is difficult to find a meaningful $\lambda$ parameter, i.e., the parameter defining the cost for moving mass between causes of death. 

The empirical analysis suggests that the discrepancy between the two measures is usually very small. Nevertheless, we do find cases where they diverge. For example, England \& Wales and Iceland display very similar mean ages at death in 1849, yet the $W_1$ Wasserstein distance indicates that their age-at-death distributions differ substantially. This arises because the $W_1$ distance reflects the \emph{absolute} differences between survivorship functions, whereas the difference in $e_0$ reflects a \emph{net difference}, where positive and negative deviations in $l(x)$ can cancel each other out. The latter is meaningful when the focus is on expected life years, but it does not necessarily capture how different the underlying age-at-death distributions are. For this reason, it is worthwhile to calculate both measures.

In many empirical applications, we can expect survivorship functions not to crossover. For instance, when comparing women and men, or populations with high versus low socioeconomic status. In such cases, differences in $e_0$ can be directly interpreted as distributional differences. The Wasserstein distance provides a particularly elegant and intuitive way to interpret differences in age-at-death distributions, linking them directly to the framework of optimal transport.

Finally, it is worth noting that other measures, such as the non-overlap index or the Kullback–Leibler divergence, can also be used to assess distributional differences. Our analysis shows that these measures are highly correlated with the $W_1$ Wasserstein distance and yield broadly similar conclusions. Why is this strong association so evident in the demographic setting of age-at-death distributions? Age-at-death distributions show a characteristic smooth and unimodal shape and tend to shift toward older ages as mortality declines. Although the shape is not strictly invariant, changes are typically dominated by location shifts rather than substantial alterations in dispersion or modality. Under the approximation that two such distributions differ primarily by a shift, the $W_1$ distance has a direct mathematical relationship with the non-overlap index. In this setting, both measures are effectively driven by the same underlying displacement parameter, providing a formal explanation for their high empirical correlation. Nonetheless, both measures should not be treated as interchangeable. The $W_1$ decomposition reveals that in more recent periods, differences in life table age-at-death distributions are mostly driven by shifts rather than by dispersion. Further, in most cases, the $W_1$ Wasserstein distance corresponds to the difference in life expectancy at birth. This enables researcher to use the decomposition technique to examine $e_0$ differences over time or between populations.

\section{Acknowledgements}
A slightly different version of this paper was submitted to Demographic Research Formal Relationship section. The paper was rejected as its added value was deemed very limited but there was consensus about the technical correctness. Maybe someone else finds it interesting or useful. I would like to thank Laura Ann Cilek for inspiring conversations about the Wasserstein distance and for technical discussions on its implementation for comparing life table age-at-death distributions. I am also grateful to Pavel Griogoriev, Sebastian Klüsener, Jiaxin Shi, and Wen Su for helpful comments and suggestions. I used AI for this work to assist with language editing, coding, and refining the presentation of mathematical arguments.


\end{document}